\documentclass[twocolumn]{aa}
\usepackage{graphicx}
\usepackage{natbib}
\bibpunct{(}{)}{;}{a}{}{,}

\newcommand{\kms}{\mbox{$\mathrm{km~s^{-1}}$}}

\newcommand{\Ha}{\mbox{${\mathrm H\alpha}$}}

\newcommand{\hs}{HS\,0218}

\begin{document}
\title{An evolved donor star in the long-period cataclysmic variable \object{HS\,0218+3229} \thanks{Based in part on observations collected at the Centro Astron\'omico Hispano Alem\'an (CAHA) at Calar Alto, operated jointly by the Max-Planck Institut f\"ur Astronomie and the Instituto de Astrof\'\i sica de Andaluc\'\i a (CSIC). Observations were also obtained at the FLWO Observatory, a facility of the Smithsonian Institution.}}

\author{P. Rodr\'\i guez-Gil\inst{1,2,3} \and
	M. A. P. Torres\inst{4} \and
	B. T. G\"ansicke\inst{3} \and
	T. Mu\~noz-Darias\inst{2} \and
	D. Steeghs\inst{3,4} \and
	R. Schwarz\inst{5} \and
        A. Rau\inst{6} \and
        H.-J. Hagen\inst{6}
}

\offprints{P. Rodr\'\i guez-Gil,\\\email{prguez@ing.iac.es}}

\institute{
Isaac Newton Group of Telescopes, Apartado de correos 321, E-38700, Santa Cruz de La Palma, Spain
\and
Instituto de Astrof\'\i sica de Canarias, V\'\i a L\'actea, s/n, La Laguna, E-38205 Santa Cruz de Tenerife, Spain
\and   
Department of Physics, University of Warwick, Coventry CV4 7AL, UK
\and
Harvard-Smithsonian Center for Astrophysics, 60 Garden St, Cambridge, MA 02138, USA
\and
Astrophysikalisches Institut Potsdam, An der Sternwarte 16, 14482 Potsdam, Germany
\and  
Caltech Optical Observatories, Mail Stop 105-24, California Institute of Technology, Pasadena, CA 91125, USA
\and 
Hamburger Sternwarte, Universit\"at Hamburg, Gojenbergsweg 112, 21029 Hamburg, Germany}

\date{Received 2008; accepted 2008}

\abstract{We present time-resolved spectroscopy and photometry of
HS\,0218+3229, a new long-period cataclysmic variable discovered
within the Hamburg Quasar Survey. It is one of the few systems that
allow a dynamical measurement of the masses of the stellar
components.}{We combine the analysis of time-resolved optical
spectroscopy and $R$-band photometry with the aim of measuring the
mass of the white dwarf and the donor star and the orbital
inclination.}{Cross-correlation of the spectra with K-type dwarf
templates is used to derive the radial velocity curve of the donor
star. An optimal subtraction of the broadened templates is performed
to measure the rotational broadening and constrain the spectral type
of the donor. Finally, an ellipsoidal model is fitted to the $R$-band
light curve to obtain constraints upon the orbital inclination of the
binary system.}{The orbital period of HS\,0218+3229 is found to be
$0.297229661 \pm 0.000000001$ d ($7.13351186 \pm 0.00000002$ h), and
the amplitude of the donor's radial velocity curve is $K_2 = 162.4 \pm
1.4$~\kms. Modelling the ellipsoidal light curves gives an orbital
inclination in the range $i = 59^{\mathrm{o}} \pm
3^{\mathrm{o}}$. A rotational broadening between $82.4 \pm
1.2$~\kms~and  $89.4 \pm 1.3$~\kms~is found when assuming zero and continuum
limb darkening, respectively. The secondary star has most
likely a spectral type K5 and contributes $\sim 80-85\%$ to the
$R$-band light. Our analysis yields a mass ratio of $0.52 < q <
0.65$, a white dwarf mass of $0.44 < M_1 (M_\odot) < 0.65$, and a
donor star mass of $0.23 < M_2 (M_\odot) <  0.44$.}{We find that
the donor star in HS\,0218+3229 is significantly undermassive for its
spectral type. It is therefore very likely that it has undergone
nuclear evolution prior to the onset of mass transfer.}


\titlerunning{An evolved donor star in HS\,0218+3229}
\authorrunning{P. Rodr\'\i guez-Gil et al.}

\maketitle

\section{Introduction}

Testing the current evolutionary models of the present-day Galactic population of cataclysmic variables (CVs) is not possible without a significant sample of the actual masses of the white dwarfs (i.e. the accretors) and the late-type companion stars (i.e. the donors) that form this sort of binary system. But accurate measurements of the component masses in CVs are, however, not an easy task. The donor stars are usually veiled by the extreme brightness of the accretion structures, especially the accretion disc. Furthermore, the white dwarfs are also very often hidden from view in CVs with orbital periods above the 2--3\,hour period gap, so optical studies aimed at dynamical mass measurements are scarce and usually make dubious assumptions. A common practice has been to assume that the amplitude of the radial velocity curve of the wings of the disc emission lines can be regarded as a reasonable estimate of the radial velocity amplitude of the white dwarf ($K_1$), but this assumption is usually far from being true \citep[see e.g.][]{shafteretal95-1}. Therefore, time-resolved studies of CVs in which the donor star and/or the white dwarf are exposed are chiefly important to our understanding of CV evolution.

\object{HS\,0218+3229} was discovered during our large-scale search for CVs in the Hamburg Quasar Survey (HQS), based on their spectroscopic properties or, more specifically, the presence of strong emission and/or absorption lines \citep{gaensickeetal02-2}. Fifty-three new CVs were found within this project, and the orbital period has been determined for the majority of them. The main result is that most of the HQS CVs have been found to have orbital periods above the period gap. Those include rarely outbursting dwarf novae, such as \object{GY\,Cnc} \citep[\object{HS\,0907$+$1902}, ][]{gaensickeetal00-2} or \object{RX\,J0944.5$+$0357} \citep[\object{HS\,0941$+$0411}, ][]{mennickentetal02-1}; magnetic CVs with relatively weak X-ray emission such as \object{1RXS\,J062518.2+733433} \citep[\object{HS\,0618+7336}, ][]{araujo-betancoretal03-2}, \object{RX\,J1554.2+2721} \citep[\object{HS\,1552+2730}, ][]{thorstensen+fenton02-1, gaensickeetal04-3}, and \object{HS 0943+1404} \citep{rodriguez-giletal05-2}; a number of SW\,Sextantis stars \citep{szkodyetal01-1, rodriguez-giletal04-2, rodriguez-giletal07-1}; one of the youngest pre--CVs known \citep[\object{HS\,1857$+$5144}, ][]{aungwerojwitetal07-1}; long-period CVs \citep{aungwerojwitetal05-1}; and several dwarf novae \citep{aungwerojwitetal06-1}.

This paper presents the dynamical determination of the stellar masses in the cataclysmic variable HS\,0218+3229. We present the data in Sect.~\ref{sec-obs}, and an identification spectrum in Sect.~\ref{sec-idspec}. The orbital period is measured from the optical photometry in Sect.~\ref{sec-porb} before analysing the time-resolved spectroscopic data in Sect.~\ref{sec-spec}. Modelling of the ellipsoidal modulation is carried out in Sect.~\ref{sec-ellipmod}. Finally, the system parameters of HS\,0218+3229 and the overall discussion are presented in Sect.~\ref{sec-discuss}.  

\section{Observational data\label{sec-obs}}

\subsection{Photometry}
\begin{figure}
\centerline{\includegraphics[width=6cm]{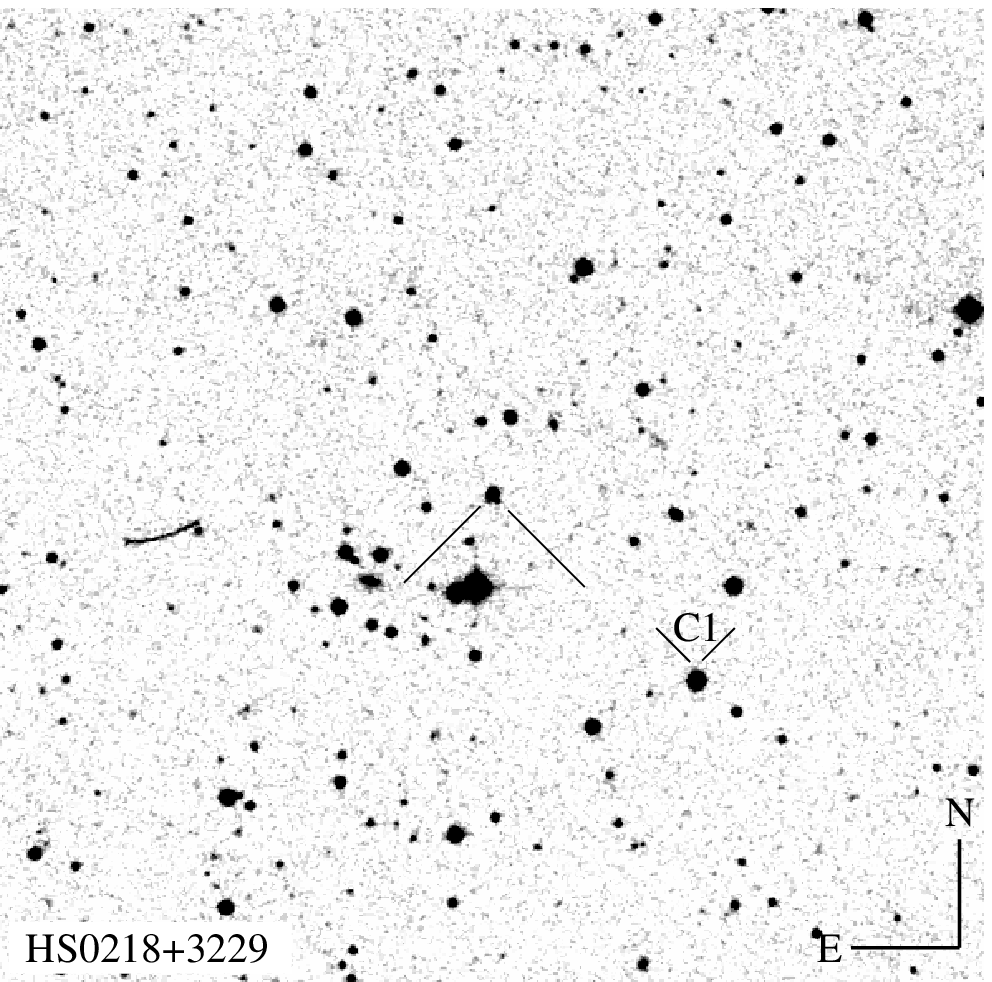}}
\caption[]{\label{fig-fc} $10\arcmin\times10\arcmin$ finding chart of
\hs~obtained from the Digitized Sky Survey~2. The
coordinates of the new CV are
$\alpha(\mathrm{J}2000)=2^\mathrm{h}21^\mathrm{m}33.50^\mathrm{s}$,
$\delta(\mathrm{J}2000)=+32\degr43\arcmin23.8\arcsec$. The star `C1' has been
used as comparison for the $R$-band differential photometry.}
\end{figure}

\paragraph{Astrophysikalisches Institut Potsdam.}
Time series $R$-band photometry of HS\,0218+3229 (hereafter \hs) was obtained on 3 nights during the period 2001 January--February using the 0.70\,m telescope of the Astrophysikalisches Institut Potsdam (AIP), Germany. The images were obtained with the camera equipped with a SITe 1024$\times$1024 pixel CCD detector and reduced in a standard way with \texttt{MIDAS}. Point spread function (PSF) photometry was done with a Perl Data Language pipeline based on the \texttt{DoPhot} package \citep{mateo+schechter89-1}. The differential magnitudes of \hs\ were derived relative to the comparison star labelled `C1' in Fig.~\ref{fig-fc}, and were converted into apparent $R$-band magnitudes using the USNO--A2.0 magnitude of the comparison star, $R_\mathrm{C1}=13.74$. The main source of uncertainty in this
conversion is the uncertainty in the USNO magnitudes, which is
typically $\simeq0.2$\,mag.

\paragraph{Fred Lawrence Whipple Observatory.}
Further time-resolved $R$-band photometry of \hs\ was secured on 2005 October 8--11 using the 1.2\,m telescope at the Fred Lawrence Whipple Observatory (FLWO) in Arizona. The images were obtained with the KEPLERCAM mosaic camera which consists of an array of four 2048$\times$2048 pixel CCD detectors. Only a small window on CCD \#4 was read out in order to minimise the dead time between images. Aperture photometry was performed on the reduced images using \texttt{IRAF}\footnote{{\texttt{IRAF}} is distributed by the National Optical Astronomy Observatories, which is operated by the Association of Universities for Research in Astronomy, Inc., under contract with the National Science Foundation.}. An optimal aperture radius of 1.5 times the full-width at half-maximum (FWHM) of the typical seeing profile was used \citep{naylor98-1}. The light curves were computed in the same manner as described for the Potsdam data. A brief summary of the observations is given in Table\,\ref{t-obslog}.

\subsection{Optical spectroscopy}

During an identification run of HQS CV candidates at the 2.2\,m Calar Alto telescope on 2000 September 20, we obtained a pair of blue/red spectra of \hs~with the CAFOS spectrograph (Table~\ref{t-obslog}). We used the B--200 and R--200 gratings in conjunction with a 2\arcsec~slit, which provided a spectral resolution of $\simeq 10$\,\AA~(FWHM). A standard reduction was performed using the CAFOS \texttt{MIDAS} quicklook package. The
detection of strong Balmer emission lines and absorption lines of a late-type star confirmed the CV nature of \hs~and encouraged us to make the follow-up observations described in what follows.

\begin{table}[t]
\caption[]{\label{t-obslog}Log of observations}
\setlength{\tabcolsep}{1.1ex}
\begin{flushleft}
\begin{tabular}{lcccc}
\hline\noalign{\smallskip}
\hline\noalign{\smallskip}
UT Date & Coverage & Filter/Grating & Exp. & \#\,Frames \\
         &   (h)   &                &   (s) & \\  
\hline\noalign{\smallskip}
\multicolumn{5}{l}{\textbf{Calar Alto 2.2\,m, CAFOS spectroscopy}} \\
2000 Sep 20    & -- &  B--200/R--200  & 600 & 1/1   \\
\hline\noalign{\smallskip}
\hline\noalign{\smallskip}
\multicolumn{5}{l}{\textbf{AIP photometry}} \\
2001 Jan 08    & 4.09 &  $R$  & 60 & 223   \\
2001 Jan 16    & 6.11 &  $R$  & 60 & 330   \\
2001 Feb 15    & 4.78 &  $R$  & 60 & 251   \\
\hline\noalign{\smallskip}
\multicolumn{5}{l}{\textbf{Calar Alto 3.5\,m, TWIN spectroscopy}} \\
2002 Oct 28    & 6.63   & T08/T01 & 600 & 38 \\
2002 Oct 29    & 3.10   & T05/T06 & 600 & 18 \\
\hline\noalign{\smallskip}
\multicolumn{5}{l}{\textbf{FLWO 1.2\,m photometry}} \\
2005 Oct 08    & 5.65 &  $R$  & 60 & 244   \\
2005 Oct 09    & 7.52 &  $R$  & 60 & 377   \\
2005 Oct 10    & 7.17 &  $R$  & 60 & 372   \\
2005 Oct 11    & 7.97 &  $R$  & 60 & 413   \\
\noalign{\smallskip}\hline
\end{tabular}
\end{flushleft}
\end{table}

Time-resolved spectroscopy of \hs~was obtained with the double-armed TWIN spectrograph on the 3.5\,m telescope in Calar Alto on 2002 October 28--29 (Table~\ref{t-obslog}). On the first night, a total of 38 blue and red spectra were acquired using the T08 and T01 gratings, respectively, and a 1.2\arcsec~slit. The wavelength ranges $\lambda\lambda3800-5560$ and $\lambda\lambda5800-7390$ were sampled at 2.2 and 1.6 \AA~resolution (FWHM), respectively. The gratings T05 and T06 and the same slit width were selected for the second night. This granted access to the ranges $\lambda\lambda3985-5050$ and $\lambda\lambda5910-6990$ at 1.1 and 1.2 \AA~resolution (FWHM) in the blue and the red, respectively. Spectra of the G191--B2B flux standard were taken to derive the instrument response function. Both the identification spectra and the spectra obtained on the first night showed a red continuum and absorption features originating in the photosphere of a mid-K secondary star. Hence, we also took spectra of six K stars of different spectral sub-type during the second night. For wavelength calibration we acquired spectra of a HeAr lamp throughout each night.

Prior to optimal extraction of the individual spectra \citep{horne86-1}, the raw images were bias-subtracted and flat-field corrected. Sky emission was removed during the optimal extraction of the spectra. The pixel--wavelength data were well fitted by a fourth-order polynomial, with an {\sl rms} always smaller than one tenth of the spectral dispersion in all cases. The wavelength scale for each spectrum was derived by interpolating between the two nearest arc spectra. The accuracy of our wavelength calibration was improved by using the night-sky lines to correct for any zero-point offset.

The reduction steps and one-dimensional spectrum extraction were performed within \texttt{IRAF}, whilst the wavelength calibration was performed in \texttt{MOLLY}\footnote{Written by T. R. Marsh,\\ http://www.warwick.ac.uk/go/trmarsh}.

\begin{figure}
\centering \includegraphics[height=7cm]{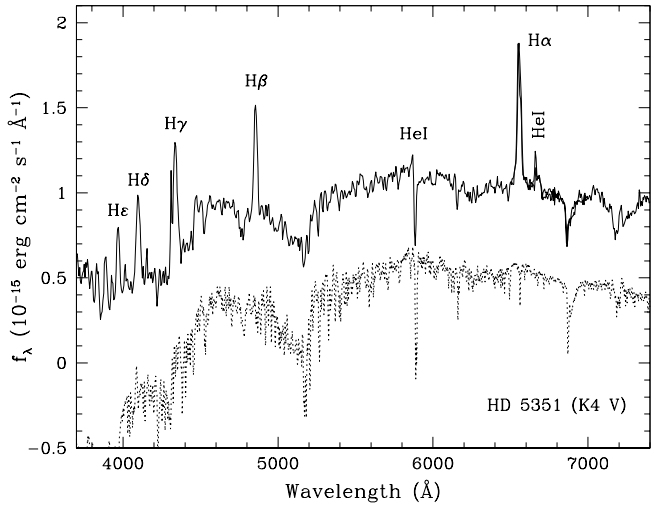}
\caption{\label{fig_idspec} Discovery spectrum of \hs~obtained with the 2.2\,m telescope at Calar Alto in September 2000 (solid line). The spectrum of a K4 V star (HD 5351) is shown as a dotted line, shifted down by 0.5 units for clarity. The template spectrum was extracted from the spectral catalogue of \cite{jacobyetal84-1}. The absorption-line spectrum of \hs\ is characteristic of a mid-K star.}
\end{figure}

\begin{figure*}[t!]
\includegraphics[width=9cm]{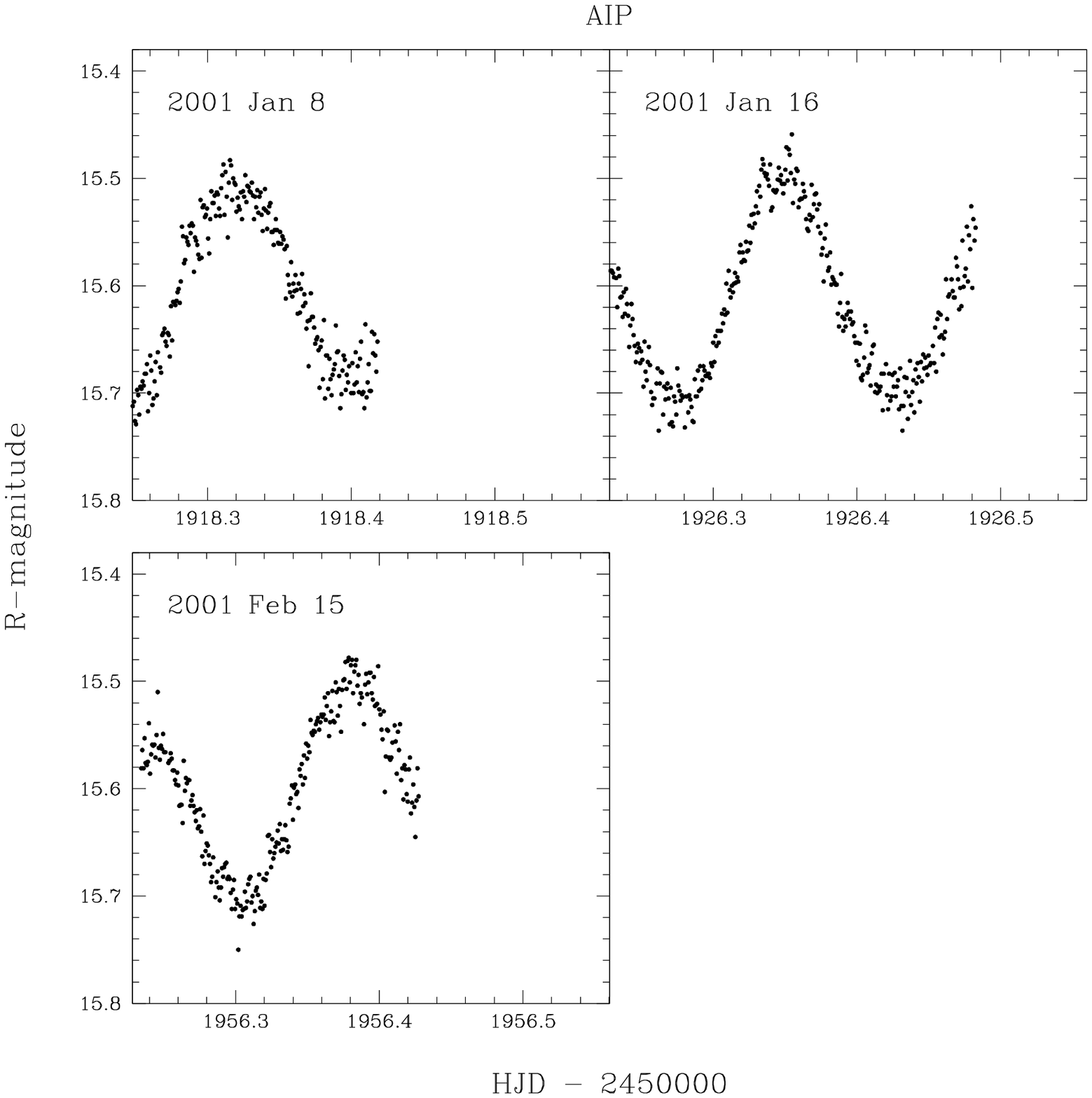}
\includegraphics[width=9cm]{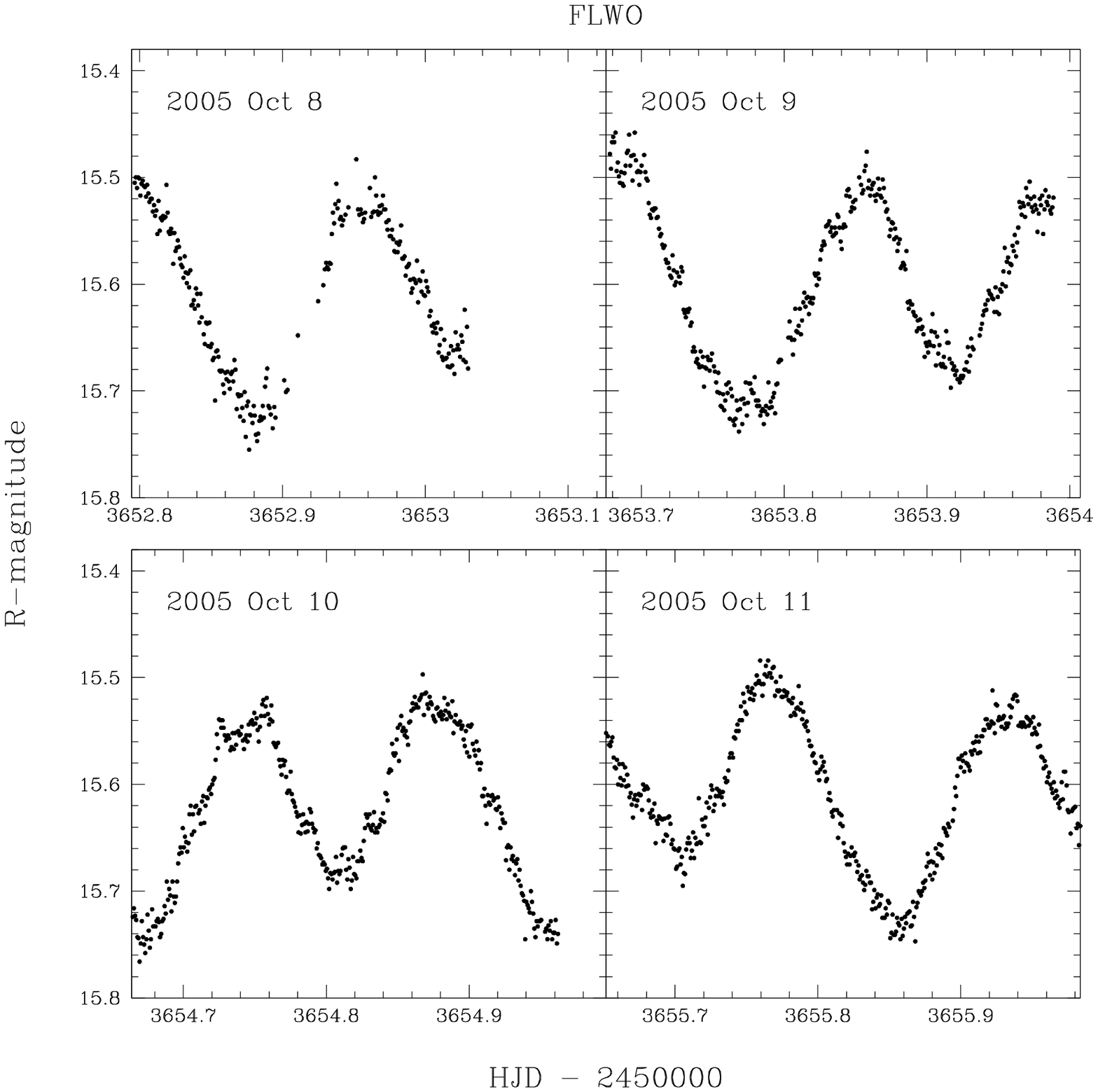}
\caption{\label{fig_lightcurves} $R$-band light curves of HS\,0218+3229 obtained at the AIP ({\em left}) and the FLWO ({\em right}).}
\end{figure*}

\section{Identification spectroscopy\label{sec-idspec}}

One of the remarkable features of the identification spectrum of \hs\
(Fig.\,\ref{fig_idspec}) is its red continuum, which reveals the dominance of the companion star. The absence of strong TiO absorption bands suggests a spectral type earlier than M. In fact, the broad absorption observed at $\sim 5200$\,\AA~ (produced by MgH at 5180 \AA, a TiO band at $\lambda\lambda4954-5200$, and a jump in the continuum due to the \ion{Mg}{i} triplet at $\lambda\lambda5168-5185$) constitutes a clear signature of a K-type star \citep[][~and references therein]{mcclintock+remillard90-1}. In Sect.~\ref{sec-spectype} we present a more accurate determination of the secondary star spectral type based on measurements of the rotational broadening of its absorption lines.


\section{The orbital period of \hs \label{sec-porb}}

The $R$-band light curves of \hs~(Fig.~\ref{fig_lightcurves}) exhibit a clear quasi-sinusoidal modulation with variable minima. Deeper and shallower minima alternate in a periodic basis every $\sim 3.5$\,h. This behaviour, together with an optical spectrum dominated by the donor star (Fig.~\ref{fig_idspec}), make us identify this variation with a classical ellipsoidal modulation resulting from the changing projected area of the Roche lobe with orbital phase. The deeper minima in the ellipsoidal light curve are caused by the stronger gravity darkening on the hemisphere facing the white dwarf, so they must occur at orbital phase 0.5. Therefore, the actual orbital period of \hs~must be twice the separation between consecutive minima, which is confirmed by the radial velocity curve of the donor star (Sect.~\ref{sec-rvs}).

\begin{table}
\caption[]{\label{table_oc} Timings of phase 0 and 0.5 (light curves only).}
\setlength{\tabcolsep}{1.1ex}
\begin{center}
\begin{tabular}{lrr}
\hline\noalign{\smallskip}
\hline\noalign{\smallskip}
~~~~~~~~~~$T_0$ & Cycle & $O-C$ \\
($\mathrm{HJD} - 2450000$) &      &   (s)~~ \\  
\hline\noalign{\smallskip}
\hline\noalign{\smallskip}
1918.39911  & $-$5836   & 239   \\
1926.27401  & $-$5809.5 & 94    \\
1926.42749  & $-$5809   & 514   \\
1956.30491  & $-$5708.5 & 1018  \\
2576.44547  & $-$3622   & $-$1496 \\
2577.33716  & $-$3619   & $-$1495 \\
3652.88176  & $-$0.5    & 156   \\
3653.77448  & 2.5     & 245   \\
3653.91820  & 3       & $-$177  \\
3654.67436  & 5.5     & 953   \\
3654.80932  & 6       & $-$226  \\
3654.95944  & 6.5     & $-$96   \\
3655.70444  & 9       & 70    \\
3655.85465  & 9.5     & 208   \\
\hline\noalign{\smallskip}
\end{tabular}
\end{center}
\end{table}

\begin{figure}
\centering \includegraphics[width=\columnwidth]{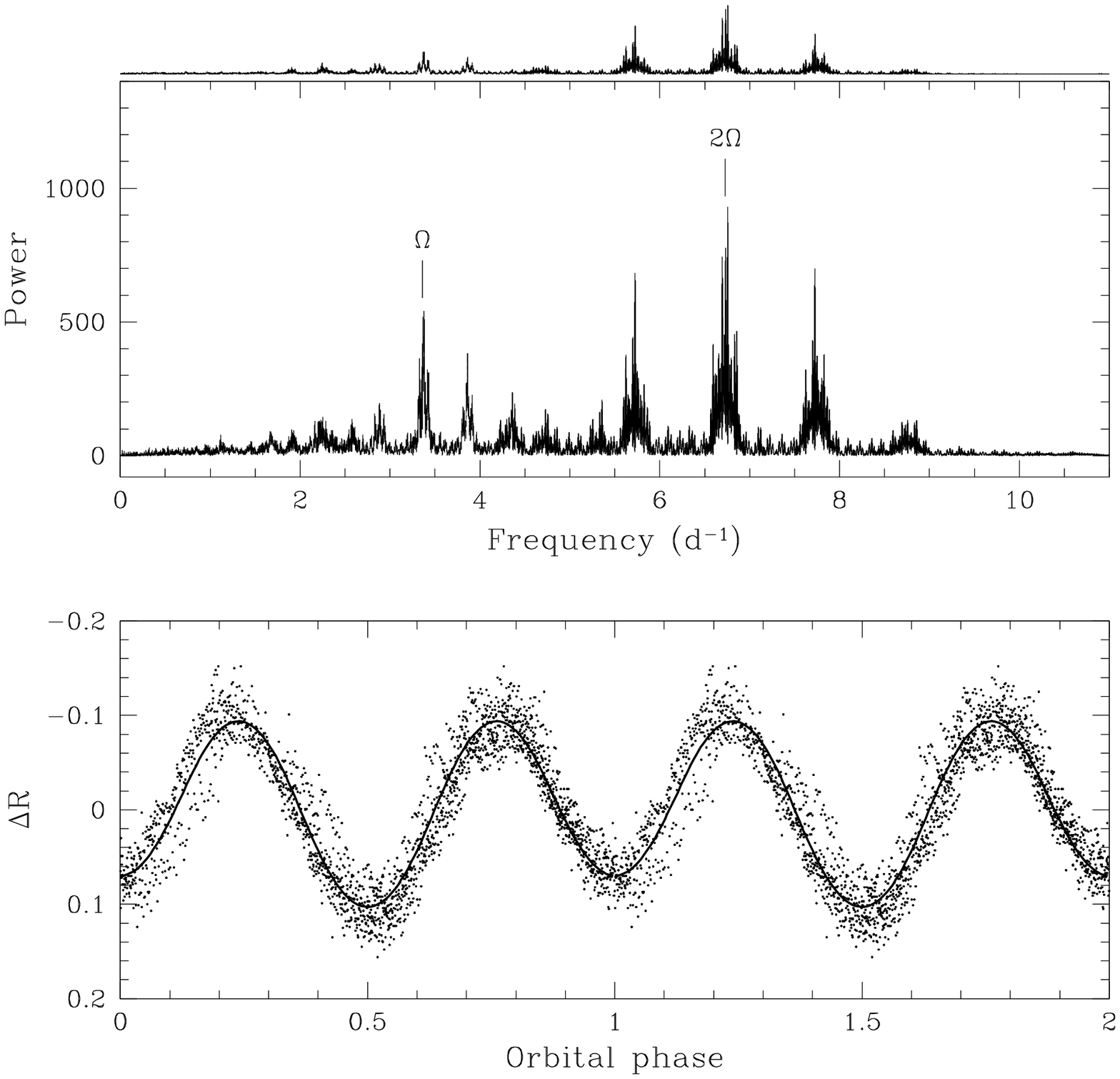}
\caption{\label{fig_aovphot} {\em Top main panel}: analysis-of-variance (AOV) periodogram computed from the $R$-band photometric data (Fig.~\ref{fig_lightcurves}). The periodogram shown on top was constructed from a fake sine wave with a frequency of $2\Omega$ sampled at the actual data points. The structure of both periodograms are nearly the same. {\em Bottom panel}: the $R$-band photometric data folded according to the ephemeris given in Eq.~\ref{eq1}. The solid line is the best ellipsoidal fit to the curve (see Sect.~\ref{sec-ellipmod}). No phase binning has been applied. A whole orbit has been duplicated for clarity.}
\end{figure}

As the photometry spans much longer than the spectroscopy, the analysis of the light curves provided a more accurate orbital period determination. The analysis-of-variance \citep[AOV, ][]{schwarzenberg-czerny89-1} periodogram computed from our light curves is presented in Fig.~\ref{fig_aovphot}. The highest peak is centred at $\nu \simeq 6.73\,\mathrm{d}^{-1}$ ($\simeq 3.57$\,h), whilst the second highest lies exactly at half that frequency ($\Omega \simeq 3.36\,\mathrm{d}^{-1}$; $P_\mathrm{orb} \simeq 7.13$\,h), which is the actual orbital frequency.

In order to improve the accuracy of the orbital period we calculated a linear ephemeris from the times of all available photometric minima derived from Gaussian fits. Additionally, two measurements of the instant of zero phase (i.e. inferior cojunction of the secondary star, $T_0$) were obtained from sine fits to the secondary star's radial velocity curves presented in Sect.~\ref{sec-rvs}, which were also included in the calculations. The measurements of the times of phase 0 (and also 0.5 for the ellipsoidal light curves) are presented in Table~\ref{table_oc}. The resulting orbital ephemeris is:

\begin{equation}
\label{eq1}
T_0(\mathrm{HJD}) = 2\,453\,653.028599(3) + 0.297229661(1) \times E~,
\end{equation}      

\noindent
where the numbers in parentheses quote the uncertainty in the last digit.

Figure~\ref{fig_aovphot} also shows all the $R$-band data folded on the orbital period using the ephemeris given in Eq.~\ref{eq1}. The ellipsoidal modulation is apparent. The shape of the folded light curves, however, shows significant deviations between the AIP and FLWO data. The 2001 January 8 light curve clearly displays the phase-0.5 minimum and phase-0.75 maximum shifted by $\sim -0.05$ cycle with respect to the FLWO light curve. This cannot be the effect of an inaccurate orbital period determination as the phase-0 minima and phase $0.75-0$ branches match very well in both sets of data. In addition, the 2001 January 16 AIP light curve exhibits a broader phase-0 minimum, with the phase $0-0.25$ rising branch lagging that of the FLWO and almost equal primary and secondary minima. This behaviour indicates that the ellipsoidal modulation is actually contaminated by other sources of variation, likely originating in the accretion disc and/or the donor star itself (e.g. irradiation).

On the other hand, the $R$-band light curves of \hs~also display higher frequency activity (see e.g. the 2005 October 10 light curve in Fig.~\ref{fig_lightcurves}). To analyse these variations we first subtracted the ellipsoidal modulation (by means of a smoothed, Fourier series-fitted version of the curves) and then computed an AOV periodogram from all the detrended light curves. The combined power spectrum does not show any predominant frequency. However, the periodogram computed from the 2005 October 9 data alone (not shown) exhibits a prominent peak at $\nu \simeq 37.4\,\mathrm{d}^{-1}$, which corresponds to a period of $P \simeq 38.9$\,min ($= 0.027$\,d), approximately 10\% of the orbital period. The periodograms calculated from the other individual light curves (with the exception of 2005 October 8) exhibit their strongest peaks at close but lower frequencies. This might suggest the presence of quasi-periodic oscillations in \hs, but a much more longer light curve coverage would be needed to draw any firm conclusions.    


\section{Time-resolved spectroscopy\label{sec-spec}}

\subsection{The radial velocity amplitude of the donor star\label{sec-rvs}}

The radial velocity curve of the secondary star in \hs\ was measured
from the red spectra using the method of cross-correlation with the
spectrum of a template star \citep{tonry+davis79-1}. Five template
stars with spectral types ranging from K0\,V to K5\,V were acquired on
the night of 2002 October 29 to extract the radial velocities. Prior to
the cross-correlation, the spectra were re-sampled onto a common
logarithmic wavelength scale and normalised by means of a low-order
spline fit to the continuum. Individual velocities were then extracted
by cross-correlation with each template star in the range
$\lambda\lambda6090-6520$. We chose this spectral interval because it
contains lines useful not only for radial velocity measurements, but
also for spectral-type classification and rotational broadening
measurements (see Sect.~\ref{sec-spectype}).

A first look at the velocities indicates a period of $\simeq 0.3$\,d,
which confirms the ellipsoidal nature of the $R$-band light curves
(Sect.~\ref{sec-porb}). We therefore performed least-squares sine fits
to the radial velocity data with the period fixed at the photometric
orbital period. The resulting fit parameters are summarised in
Table~\ref{tab-rvfits}. Although all the values are very similar, we adopt the
K5\,V parameters of the radial velocity curve since the spectral type of the
donor star is shown to be most likely K5 (see Sect.~\ref{sec-spectype}):

\begin{itemize}

\item[] $K_{2} = 162.4 \pm 1.4$ km s$^{-1}$

\item[] $\gamma_2 = -54.2 \pm 2.9 $ km s$^{-1}$


\end{itemize}

\begin{figure}
\centering
\includegraphics[width=\columnwidth]{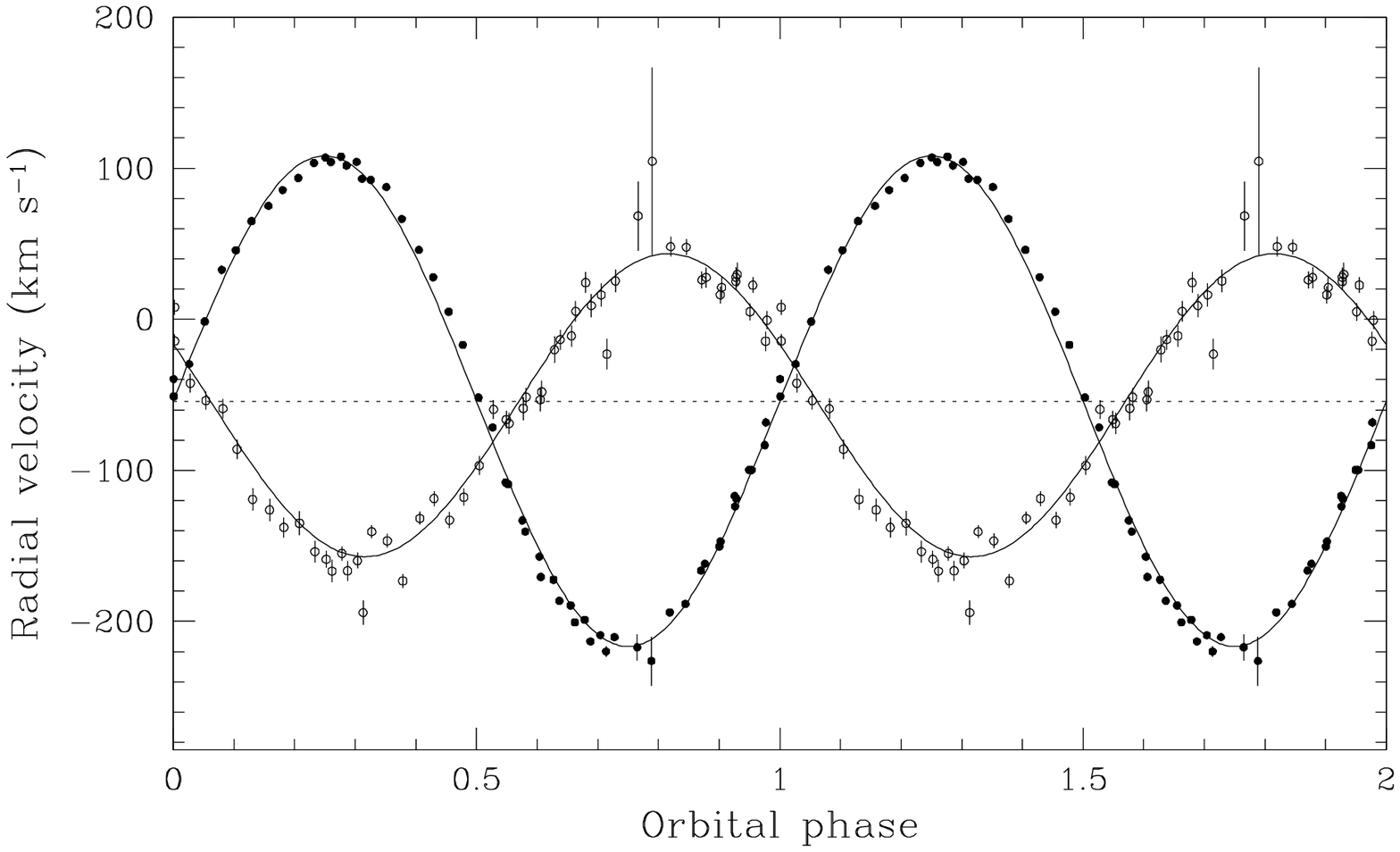}
\caption{\label{fig-vel} Radial velocity curves of the secondary star (solid dots) and the \Ha~line wings (open circles) in \hs. The radial velocities of the donor star were extracted by cross-correlation with a K5\,V template star. Solid curves are the best sine fits to the radial velocities. The horizontal dashed line marks $\gamma_2 = -54.2$ \kms. The orbital cycle has been plotted twice.}
\end{figure}

\noindent All quoted uncertainties are
1-$\sigma$. The phase-folded radial velocity curve is shown in
Fig.~\ref{fig-vel}.

\begin{table*}
\caption{Radial velocity parameters (errors at 1$\sigma$). \label{tab-rvfits}}
\label{tablerv}
\begin{center}
\begin{tabular}{ccccccc}
\hline\noalign{\smallskip}
\hline\noalign{\smallskip}
Template &Spectral &$\gamma$      &$\gamma_{2} ~ ^{(a)}$       &$K_{_2}$         &$T_0$                & $\chi^2_{\nu}$\\
         &type     &(km s$^{-1}$) &(km s$^{-1}$)      &(km s$^{-1}$)    & $(\mathrm{HJD}-2452576)$          & (d.o.f=53)     \\
\hline\noalign{\smallskip}
\hline\noalign{\smallskip}

HD 10780    & K0\,V & $-56.2 \pm 1.1$ & $ -53.5 \pm 1.2$ &161.6 $\pm$ 1.5  & 0.4457 $\pm$ 5$\times$10$^{-4}$ & 9.1  \\
HR 10476    & K1\,V & $-20.9 \pm 1.1$ & $ -54.7 \pm 1.1$ &161.7 $\pm$ 1.5  & 0.4455 $\pm$ 5$\times$10$^{-4}$ & 9.4  \\
HR 16160    & K3\,V & $-76.4 \pm 1.0$ & $ -50.7 \pm 1.2$ &162.3 $\pm$ 1.4  & 0.4457 $\pm$ 4$\times$10$^{-4}$ & 10.1 \\
HD 6660     & K4\,V & $-59.7 \pm 1.0$ & $ -54.9 \pm 1.6$ &162.2 $\pm$ 1.4  & 0.4456 $\pm$ 4$\times$10$^{-4}$ & 13.5  \\
HD 10436    & K5\,V & $ -2.8 \pm 1.0$ & $ -54.2 \pm 2.9$ &162.4 $\pm$ 1.4  & 0.4457 $\pm$ 4$\times$10$^{-4}$ & 11.9  \\
\hline\noalign{\smallskip}

\end{tabular}
\begin{tabular}{c}
\\
$^{(a)}$ $\gamma_{_2}$: Systemic velocity after adding the radial velocity of the template star as found in Gontcharov (2006).
\end{tabular}
\end{center}
\end{table*}

\begin{table*}
\begin{center}
\caption{Spectral classification and rotational broadening. \label{tab-chi}}
\begin{tabular}{lccccccc}
\hline\noalign{\smallskip}
\hline\noalign{\smallskip}
Template &Spectral & $v \sin i ~ ^{(a)}$ (km s$^{-1}$) & $f$  &$\chi^2_{\nu}$ &  $v \sin i ~ ^{(b)}$ (km s$^{-1}$)  \\
 	 &type	   & ($\mu=0.5$)              &      &(d.o.f.=496)   &  [$\mu=0.0$, $\mu$ at continuum] \\ 
\hline\noalign{\smallskip}
\hline\noalign{\smallskip}
 
HD10780  & K0~V	   & 89.5 $\pm$ 1.2 & 1.20 $\pm$ 0.01   & 2.1 & [84.7 $\pm$ 1.8, 91.4 $\pm$ 2.0] \\ 
HD10476  & K1~V	   & 89.6 $\pm$ 1.2 & 1.17 $\pm$ 0.01   & 1.8 & [84.8 $\pm$ 1.6, 91.2 $\pm$ 4.3] \\
HD16160  & K3~V	   & 89.0 $\pm$ 1.2 & 0.97 $\pm$ 0.01   & 1.0 & [84.3 $\pm$ 1.2, 91.1 $\pm$ 1.3] \\ 
HD 6660   & K4~V	   & 87.7 $\pm$ 1.2 & 0.728 $\pm$ 0.007 & 1.0 & [83.2 $\pm$ 1.1, 90.0 $\pm$ 1.1] \\
HD10436  & K5~V	   & 86.9 $\pm$ 0.6 & 0.785 $\pm$ 0.008 & 0.9 & [82.4 $\pm$ 1.2, 89.4 $\pm$ 1.3] \\

\hline\noalign{\smallskip}

\end{tabular}
\begin{tabular}{l}
$^{(a)}$ Uncertainties are 1-$\sigma$ corresponding to $\chi^{2}_{\rm min}+1$ (Lampton, Margon \& Bowyer 1976). \\
$^{(b)}$ Uncertainties are 1-$\sigma$ and were estimated using the Monte-Carlo method by simulating a total of 1000000 \\
copies of our target spectrum using the bootstrap technique outlined in \cite{steeghs+jonker07-1}. \\
\end{tabular}
\end{center}
\end{table*}     

\subsection{Spectral type and rotational broadening\label{sec-spectype}}

To determine the spectral type of the donor star and the rotational
broadening of its absorption lines, we used the technique outlined in
\cite{marshetal94-1}. It is based on the search for the lowest
residual obtained when subtracting a set of templates from the
Doppler-corrected, average spectrum of the target.

We proceeded as follows: we used the spectra acquired during the
second night, as both target and template spectra were acquired with
the same instrumental setup. This is essential to derive a reliable
value for the rotational broadening ($v \sin i$), in particular when $v \sin i$ is
of the same order as the instrumental resolution. First, the 18
target spectra were Doppler-corrected to the rest frame of the donor
star by using the radial velocity parameters derived in
Sect.~\ref{sec-rvs}. Next, we produced an average spectrum after
assigning different weights to the individual spectra to maximise the
signal-to-noise ratio of the sum. The template spectra were broadened from
70 to 110 km s$^{-1}$ in steps of 1.0 km s$^{-1}$ through convolution
with the spherical rotational profile of Gray (1992) with a linearised
limb-darkening coefficient of $0.5$. Each broadened version of a
template spectrum was scaled by a factor $f$ (representing the
fractional contribution of light from the secondary star) and
subtracted from the target Doppler-corrected average. Then a $\chi^2$
test on the residuals was performed in the range
$\lambda\lambda6090-6520$ and the optimal values of $f$ and $v \sin i$
were provided by minimising $\chi^{2}$.

Besides the instrumental
broadening, the secondary star's photospheric lines are also smeared
due to the orbital motion of the donor star during the length of the
exposure. With 600-s exposures, the orbital smearing is $\le t_{\rm
exp}{2 \pi\over P} {K_2} = 24$ km s$^{-1}$. This effect can be
corrected producing a spectral template by averaging 18 smeared copies
of a template spectrum with the same weights as used to average the
spectra of the target. We performed this correction, finding no
significant difference in the values calculated for the rotational
broadening.

The minimisation of $\chi^2_\nu$ (Table~\ref{tab-chi}) shows that the
spectral type of the secondary star in \hs\ is later than K1. Also, a K0-1 V donor star is rejected as the templates give $f >
1$. Similar results for the spectral classification were obtained
when repeating the above procedure on the 38 target spectra acquired
during the first night. Additional constraints on the spectral
type of the donor star are provided by the infrared colours of \hs\
after correcting for reddening and assuming the same $f_{disc}$ at
each infrared bandpass. Using 2MASS magnitudes we derive
$(J-H,\,H-K,\,J-K)_0=(0.63,\,0.09,\,0.72)$ based on $\mathrm{E}(B-V)=0.071$ mag \citep{schlegeletal98-1}. The uncertainties in the infrared
colours are 0.04 mag. The above colours reject an M-type dwarf $(J-K > 0.86,\,H-K > 0.165)$ and support a K4--K5 dwarf \citep[see e.g.][]{bessell+brett88-1}. For comparison, a K5 dwarf has colours of
$(0.61,\,0.11,\,0.72)$. Further visual inspection of the average spectra
and templates supports a K5 spectral type based on the strength
of the weak TiO bands in the spectrum and by comparing the relative
intensities of the absorption lines in the interval $\lambda\lambda
6400 - 6530$. Hereafter, in accordance with the above analysis, we adopt a K5 spectral type for the donor star.

A rotational broadening measurement of $86.9 \pm 0.6$ km s$^{-1}$ was
obtained from the K5\,V template. However, the value of the
limb-darkening coefficient for the absorption lines is unknown and
this limits our ability in evaluating $v \sin i$. Limb-darkening
coefficients for the absoption lines in late-type stars are expected
to be smaller than those for the continuum (Collins \& Truax
1995). Therefore, we repeated the above analysis for both a null
linearised limb-darkening coefficient and for the value expected for
the continuum. The latter was selected for each template from \cite{claretetal95-1}.  We provide the
results for each template in Table 3 (column 6). As one can see, our
uncertainty in the limb-darkening leads to values of $v \sin i$ of
$82.4-89.4$~km s$^{-1}$ (K5\,V template).

\subsection{The radial velocity amplitude of the white dwarf\label{sec-K1}}

We measured the radial velocity curve of the \Ha~line wings by using
the double-Gaussian technique of \cite{schneider+young80-2}. In an
ideal world, with an axisymmetric accretion disc dominating line
emission, the line wings would provide a good estimate of $K_1$. We
obtained radial velocity curves of the \Ha~wings for a number of
Gaussian ($\mathrm{FWHM} = 200$\,\kms) separations between 600 and
2900\,\kms~in 100\,\kms~steps. The curve for a 1300\,\kms~separation
gave the best results and is shown in Fig.~\ref{fig-vel}. The wings are clearly delayed
by $\sim 0.1$ cycle, but their $\gamma$-velocity of $-57.6 \pm
0.8$\,\kms~ (derived from a sine fit) is very similar to that of the
donor star ($\gamma_2 = -54.2 \pm 1.4$\,\kms). The sine fit gives an
amplitude of the wings radial velocity curve of $100.3 \pm 1.2$\,\kms,
which provides a mass ratio of $q=K_1/K_2=0.618 \pm 0.009$ if assumed
to be the radial velocity amplitude of the white dwarf. This mass
ratio is equal to the measured from the rotational broadening of the
donor star absorption lines. Despite the phase shift that the \Ha\ radial velocities show (Fig.~\ref{fig-vel}), the measured $K_1$ value seems to be reliable in this case.


\section{Ellipsoidal light curve modelling\label{sec-ellipmod}}
To model the ellipsoidal variation we used a binary code which uses full Roche geometry and a black body model to account for the emission of the distorted companion. In order to carry out the numerical integration the donor star surface is divided in small triangular tiles of equal area which cover completely the Roche-lobe filling star. 
In this model, the disc contribution to the total light ($f_\mathrm{disc}=1-f$) is independent of the orbital phase and is scaled to a certain factor of the mean flux from the companion. The limb darkening coefficients are obtained from \cite{al-naimiy78-1} and the $\beta$ gravity darkening exponent is fixed at 0.08 \citep{lucy67-1}. According to \cite{gray92-1} the effective temperature of a K5\,V star is 4557~K, which was adopted as the polar temperature. The synthetic ellipsoidal modulation should only depend on the orbital inclination angle, the mass ratio ($q = M_{2}/M_{1}$, where $M_{2}$ and $M_{1}$ are the masses of the donor star and the compact object, respectively), and the relative disc contribution to the light curve ($1-f$). 
In Sect~\ref{sec-spectype} we obtained  $1-f \sim 0.22$ by measuring the absorption lines in the range $\lambda\lambda6090-6520$. This spectral region is contained in the $R$ band and we can therefore use this value in our fits. However, we think that the contamination in the shape of the light curve may be due to night-to-night variations of the disc contribution (see Sect.~\ref{sec-porb}). Hence, the value obtained during the spectroscopic campaign may not be the correct one for the photometric data.

In order to establish a conservative lower limit to the inclination angle ($i$) of the system we have fitted the light curve by using $1-f = 0$, obtaining $i \ge 50\degr$. Since the light curves do not show eclipses we find an upper limit of $i \le 62\degr$ from elemental trigonometry involving the maximum disc radius allowed by tidal interaction with the donor star \citep{paczynski77-1} and the Roche lobe radius of the latter. Thus, we obtain $50^{\circ} \leq i \leq 62^{\circ}$ independently of $f_\mathrm{disc}$. The best fit to the ellipsoidal light curve was finally found for $f_\mathrm{disc}=0.15$, $q=0.62$, and $i=59\pm3^{\circ}$. We have overplotted this synthetic model to the actual folded ellipsoidal curve in Fig.~\ref{fig_aovphot}.

\section{Discussion\label{sec-discuss}}
\subsection{Parameter estimates}

Assuming that the companion star is synchronised with the binary
motion and fills its Roche lobe, it is possible to calculate the mass
ratio through the expression $ v \sin i = 0.462~K_{2}~q^{1/3} (1 +
q)^{2/3} $ (e.g. Wade \& Horne 1988). Thus we derive $0.52 < q < 0.65$
from $K_2 = 162.4 \pm 1.4$ km s$^{-1}$ and the 99\% confidence
range for the rotational broadening ($82.4-2.33\sigma$~km s$^{-1}$ $<
v \sin i <$ $89.4+2.33\sigma$ ~km s$^{-1}$). This mass-ratio range
implies a primary velocity amplitude of 84~km s$^{-1}$ $< {K_1}=q{K_2}
<$ 106~ km s$^{-1}$. The value of $100.3 \pm 1.2$\,\kms~measured from
the \Ha~line wings lies almost in the middle of this interval.

The velocity amplitude and orbital period imply a mass function of
$f({M_1})=  P K_2^3 / 2 \pi \mathrm{G}$ = $0.132 \pm 0.001$~M$_\odot$
= $M_1 \sin^3 i / (1+q)^2$. Combining the values for $f({M_1})$ and
our extreme limits on $q$ and $i$, we constrain the masses of the
compact object and the companion star to be in the range: $0.44 < M_1
< 0.65$ M$_\odot$ and $0.23 < M_2 < 0.44$
M$_\odot$. Table~\ref{tablesp} gives the determined system parameters
for the extreme values of the inclination given by the ellipsoidal
modelling. The uncertainties in the masses for a given inclination are
dominated by the uncertainty in the value of $v \sin i$. For
comparison, using the resulting value obtained for a limb-darkening
coefficient of 0.5 ($v \sin i = 86.9 \pm 0.6$ km s$^{-1}$),  we
derive: $q= 0.61 \pm 0.01$, $M_{1}=0.54 \pm 0.03$ M$_\odot$,
$M_{2}=0.33 \pm 0.02$ M$_\odot$ and $K_{1}=99 \pm 1$~km s$^{-1}$. To
determine the above 1-$\sigma$ uncertainties a Monte Carlo
approach was used in which we draw one million random values for the
observed quantities. We treated $K_{2}$ and $v \sin i $ as being
normally distributed about their measured values with standard
deviations equal to the uncertainties on the measurements. For $i$ the
distribution was taken to be uniform. Given the stellar masses
and orbital period, we can constrain the semi-major axis ($a$) of the
orbit using Kepler's Third Law, the radius of the secondary star
($R_2$) using Eggleton's (\citeyear{eggleton83-1}) expression for the
effective radius of the secondary Roche Lobe, and the luminosity of
the secondary star ($L_2$) using Stefan-Boltzmann's Law and the 4557 K
effective temperature inferred from the spectral type \citep{gray92-1}. These quantities are listed in Table~\ref{tablesp}.

To constrain the distance to \hs\ we have adopted the luminosity of the donor star given in Table~\ref{tablesp}, which, combined with the bolometric correction and the $(V-R)$ colour of a K5 V star, yield an absolute magnitude of $M_R=5.96-5.65$. We have calculated a dereddened magnitude of the secondary star of $R=15.65-15.72$ using the apparent magnitude of \hs\ ($R=15.6$), $A_R=0.188$ and a $\sim 15-20$\% veiling from the accretion disc (Sect.~\ref{sec-ellipmod}). Therefore, the distance modulus relation provides a distance to \hs\ of $0.87-1.0$ kpc. Note that this distance estimate must be considered as approximate given the uncertainties in our photometric calibration based on USNO A2.0 magnitudes of field stars.

\begin{table}
\caption{Summary of derived system parameters.}
\label{tablesp}
\begin{center}
\begin{tabular}{ccc}
\hline\noalign{\smallskip}
\hline\noalign{\smallskip}
Parameter       & $i=56^{\circ}$  & $i=62^{\circ}$     \\
\\
\hline\noalign{\smallskip}
\hline\noalign{\smallskip}
$M_1$ (M$_\odot$) & $0.54-0.65$ & $0.44-0.54$ \\
$M_2$ (M$_\odot$) & $0.28-0.44$ & $0.23-0.36$ \\
$R_2$ (R$_\odot$) & $0.57-0.67$ & $0.53-0.63$ \\
$a$ (R$_\odot$)   & $1.76-1.93$ & $1.65-1.81$ \\
$L_2$ (L$_\odot$) & $0.12-0.17$ & $0.11-0.15$ \\
\hline\noalign{\smallskip}
\end{tabular}
\end{center}
\end{table}

\subsection{An evolved donor star}

\begin{figure}
\centering
\includegraphics[width=\columnwidth]{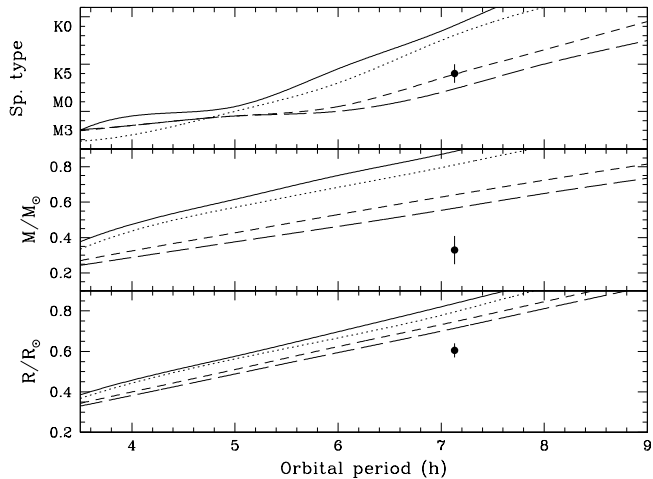}
\caption{\label{fig-evolseq} Spectral type, mass, and radius as a function of orbital period for four different evolutionary sequences with different initial central hydrogen abundance ($X_\mathrm{c}$) at the onset of mass transfer. The initial donor mass is $M_2=1.0$~M$_\odot$ and the mass transfer rate is $\dot{M}=1.5 \times 10^{-9}$~M$_\odot$~yr$^{-1}$. Solid line: $X_\mathrm{c}=0.7$ (initially unevolved donor); dotted line: $X_\mathrm{c}=0.55$; dashed line: $X_\mathrm{c}=0.16$; long-dashed line: $X_\mathrm{c}=0.05$. The location of \hs~is indicated by the large black dot. Evolutionary sequences adapted from \citet{baraffe+kolb00-1}.}
\end{figure}

As we have demonstrated, the companion star in \hs\ is most likely of
spectral type K5. A main-sequence K5 star has a nominal mass of
$0.68$~M$_\odot$ \citep{gray92-1}. The disagreement with the mass
constraints obtained above is easily explained if the secondary star
is undermassive for its spectral type, that is, has undergone nuclear
evolution {\em before} the onset of mass transfer. In fact, with a mass of
$\sim 0.33$~M$_\odot$ its nuclear timescale is much greater than the
Hubble time. Previous work on donor stars in long-period CVs
\citep[see e.g. figure 5 in][]{beuermannetal98-1} shows that most of
them at long orbital periods are underluminous, and likely evolved. In
Fig.~\ref{fig-evolseq} we plot evolutionary sequences for an initial
donor star mass of 1.0~M$_\odot$ and mass transfer onset at several
central hydrogen abundances \citep{baraffe+kolb00-1}. Note that this
is only for illustrative purposes and the actual initial donor mass in
\hs~can be different. The mass of the donor star in \hs~is much lower
than predicted by any of the sequences considered.
 
\cite{schenker+king02-1} show the evolution of the donor mean density
as a function of its mass (their figure 2) for both the ZAMS and
evolved cases. With an orbital period of 7.13 h, \hs~has a mean
density of 2.1 g cm$^{-3}$, which results in a donor mass of $\sim
0.3$~M$_\odot$ for the case of strong thermal-timescale mass
transfer. Moreover, according to \cite{schenkeretal02-1} (their figure
6), a donor star of spectral type K5--K7 is expected for a mass of
$\sim 0.3$~M$_\odot$. Both predictions for the case of an evolved
donor agree well with our measurements.

The binary system, therefore, formed with a companion star more massive than the white dwarf. In this scenario ($q=M_2/M_1 > 1$), and as a consequence of mass transfer towards the white dwarf, the Roche lobe of the donor star (initially the most massive component) shrinks faster than the thermal equilibrium radius. The secondary star thus tends to expand beyond the Roche lobe to reach the thermal equilibrium radius, which triggers an episode of thermal-timescale mass transfer until the mass ratio reverses \citep[see e.g.][]{kingetal01-1,schenkeretal02-1}. Beyond this point, normal mass transfer starts due to either angular momentum loss (i.e. secondary Roche lobe shrinkage) of nuclear evolution of the donor star (i.e. increase of the thermal equilibrium radius) and the system becomes a normal CV. The fact that the measured white dwarf mass ($M_1 \sim 0.54$~M$_\odot$) is lower than the average value for CVs \citep{smith+dhillon98-1} indicates that the transferred donor material during the thermal-timescale stage has been blown off the binary system.

Nuclear evolution of the donor star prior to normal CV life should reflect in the chemical composition of the accreted material. Because of CNO processing of the material, carbon must be significantly depleted and nitrogen subsequently enriched. Unusually low C\,{\sc iv}/N\,{\sc v} ultraviolet line flux ratios have been observed in a number of CVs \citep{gaensickeetal03-1}, suggesting that a significant fraction of the present-day CV population is descendant from systems that have passed through a thermal-timescale mass transfer phase. \hs~is therefore a promising target for ultraviolet spectroscopy to check whether N enrichment has taken place.

\begin{acknowledgements}
This work was supported by XMM-Newton Grant NNG05GJ22G. The HQS was supported by the Deutsche Forschungsgemeinschaft through grants Re 353/11 and Re 353/22. The use of the MOLLY package developed and maintained by Tom Marsh is acknowledged. Thanks to the anonymous referee for valuable input.
\end{acknowledgements}

\bibliographystyle{aa}
\bibliography{aamnem99,aabib}

\begin{thebibliography}{40}
\expandafter\ifx\csname natexlab\endcsname\relax\def\natexlab#1{#1}\fi

\bibitem[{{Al-Naimiy}(1978)}]{al-naimiy78-1}
{Al-Naimiy}, H.~M. 1978, Ap\&SS, 53, 181

\bibitem[{{Araujo-Betancor} {et~al.}(2003){Araujo-Betancor}, {G{\" a}nsicke},
  {Hagen}, {Rodr\'\i guez-Gil}, \& {Engels}}]{araujo-betancoretal03-2}
{Araujo-Betancor}, S., {G{\" a}nsicke}, B.~T., {Hagen}, H.-J., {Rodr\'\i
  guez-Gil}, P., \& {Engels}, D. 2003, A\&A, 406, 213

\bibitem[{{Aungwerojwit} {et~al.}(2006){Aungwerojwit}, {G{\"a}nsicke},
  {Rodr{\'{\i}}guez-Gil}, {Hagen}, {Araujo-Betancor}, {Baernbantner}, {Engels},
  {Fried}, {Harlaftis}, {Mislis}, {Nogami}, {Schmeer}, {Schwarz}, {Staude}, \&
  {Torres}}]{aungwerojwitetal06-1}
{Aungwerojwit}, A., {G{\"a}nsicke}, B.~T., {Rodr{\'{\i}}guez-Gil}, P., {et~al.}
  2006, A\&A, 455, 659

\bibitem[{{Aungwerojwit} {et~al.}(2007){Aungwerojwit}, {G{\"a}nsicke},
  {Rodr{\'{\i}}guez-Gil}, {Hagen}, {Giannakis}, {Papadimitriou}, {Allende
  Prieto}, \& {Engels}}]{aungwerojwitetal07-1}
{Aungwerojwit}, A., {G{\"a}nsicke}, B.~T., {Rodr{\'{\i}}guez-Gil}, P., {et~al.}
  2007, A\&A, 469, 297

\bibitem[{{Aungwerojwit} {et~al.}(2005){Aungwerojwit}, {G{\"a}nsicke},
  {Rodr{\'{\i}}guez-Gil}, {Hagen}, {Harlaftis}, {Papadimitriou}, {Lehto},
  {Araujo-Betancor}, {Heber}, {Fried}, {Engels}, \&
  {Katajainen}}]{aungwerojwitetal05-1}
{Aungwerojwit}, A., {G{\"a}nsicke}, B.~T., {Rodr{\'{\i}}guez-Gil}, P., {et~al.}
  2005, A\&A, 443, 995

\bibitem[{{Baraffe} \& {Kolb}(2000)}]{baraffe+kolb00-1}
{Baraffe}, I. \& {Kolb}, U. 2000, MNRAS, 318, 354

\bibitem[{{Bessell} \& {Brett}(1988)}]{bessell+brett88-1}
{Bessell}, M.~S. \& {Brett}, J.~M. 1988, PASP, 100, 1134

\bibitem[{{Beuermann} {et~al.}(1998){Beuermann}, {Baraffe}, {Kolb}, \&
  {Weichhold}}]{beuermannetal98-1}
{Beuermann}, K., {Baraffe}, I., {Kolb}, U., \& {Weichhold}, M. 1998, A\&A, 339,
  518

\bibitem[{{Claret} {et~al.}(1995){Claret}, {Diaz-Cordoves}, \&
  {Gimenez}}]{claretetal95-1}
{Claret}, A., {Diaz-Cordoves}, J., \& {Gimenez}, A. 1995, aaps, 114, 247

\bibitem[{{Eggleton}(1983)}]{eggleton83-1}
{Eggleton}, P.~P. 1983, ApJ, 268, 368

\bibitem[{{G{\" a}nsicke} {et~al.}(2003){G{\" a}nsicke}, {Szkody}, {de
  Martino}, {Beuermann}, {Long}, {Sion}, {Knigge}, {Marsh}, \&
  {Hubeny}}]{gaensickeetal03-1}
{G{\" a}nsicke}, B.~T., {Szkody}, P., {de Martino}, D., {et~al.} 2003, ApJ,
  594, 443

\bibitem[{{G\"ansicke} {et~al.}(2002{\natexlab{a}}){G\"ansicke}, {Beuermann},
  \& {Reinsch}}]{gaensickeetal02-1}
{G\"ansicke}, B.~T., {Beuermann}, K., \& {Reinsch}, K., eds.
  2002{\natexlab{a}}, The Physics of Cataclysmic Variables and Related Objects
  (ASP Conf. Ser. 261)

\bibitem[{{G\"ansicke} {et~al.}(2000){G\"ansicke}, {Fried}, {Hagen},
  {Beuermann}, {Engels}, {Hessman}, {Nogami}, \& {Reinsch}}]{gaensickeetal00-2}
{G\"ansicke}, B.~T., {Fried}, R.~E., {Hagen}, H.-J., {et~al.} 2000, A\&A, 356,
  L79

\bibitem[{{G\"ansicke} {et~al.}(2002{\natexlab{b}}){G\"ansicke}, {Hagen}, \&
  {Engels}}]{gaensickeetal02-2}
{G\"ansicke}, B.~T., {Hagen}, H.~J., \& {Engels}, D. 2002{\natexlab{b}}, in The
  Physics of Cataclysmic Variables and Related Objects, ed. B.~T. {G\"ansicke},
  K.~{Beuermann}, \& K.~{Reinsch} (ASP Conf. Ser. 261), 190--199

\bibitem[{{G{\"a}nsicke} {et~al.}(2004){G{\"a}nsicke}, {Jordan}, {Beuermann},
  {de Martino}, {Szkody}, {Marsh}, \& {Thorstensen}}]{gaensickeetal04-3}
{G{\"a}nsicke}, B.~T., {Jordan}, S., {Beuermann}, K., {et~al.} 2004, ApJ Lett.,
  613, L141

\bibitem[{{Gray}(1992)}]{gray92-1}
{Gray}, D.~F. 1992, {The Observation and Analysis of Stellar Photospheres}
  (Cambridge University Press)

\bibitem[{{Horne}(1986)}]{horne86-1}
{Horne}, K. 1986, PASP, 98, 609

\bibitem[{{Jacoby} {et~al.}(1984){Jacoby}, {Hunter}, \&
  {Christian}}]{jacobyetal84-1}
{Jacoby}, G.~H., {Hunter}, D.~A., \& {Christian}, C.~A. 1984, ApJS, 56, 257

\bibitem[{{King} {et~al.}(2001){King}, {Schenker}, {Kolb}, \&
  {Davies}}]{kingetal01-1}
{King}, A.~R., {Schenker}, K., {Kolb}, U., \& {Davies}, M.~B. 2001, MNRAS, 321,
  327

\bibitem[{{Lucy}(1967)}]{lucy67-1}
{Lucy}, L.~B. 1967, Zeitschrift fur Astrophysik, 65, 89

\bibitem[{{Marsh} {et~al.}(1994){Marsh}, {Robinson}, \& {Wood}}]{marshetal94-1}
{Marsh}, T.~R., {Robinson}, E.~L., \& {Wood}, J.~H. 1994, MNRAS, 266, 137

\bibitem[{{Mateo} \& {Schechter}(1989)}]{mateo+schechter89-1}
{Mateo}, M. \& {Schechter}, P.~L. 1989, in European Southern Observatory
  Astrophysics Symposia, ed. P.~J. {Grosb{\o}l}, F.~{Murtagh}, \& R.~H.
  {Warmels}, Vol.~31, 69

\bibitem[{{McClintock} \& {Remillard}(1990)}]{mcclintock+remillard90-1}
{McClintock}, J.~E. \& {Remillard}, R.~A. 1990, ApJ, 350, 386

\bibitem[{{Mennickent} {et~al.}(2002){Mennickent}, {Tovmassian}, {Zharikov},
  {Tappert}, {Greiner}, {G{\" a}nsicke}, \& {Fried}}]{mennickentetal02-1}
{Mennickent}, R.~E., {Tovmassian}, G., {Zharikov}, S.~V., {et~al.} 2002, A\&A,
  383, 933

\bibitem[{{Naylor}(1998)}]{naylor98-1}
{Naylor}, T. 1998, MNRAS, 296, 339

\bibitem[{{Paczy\'nski}(1977)}]{paczynski77-1}
{Paczy\'nski}, B. 1977, ApJ, 216, 822

\bibitem[{{Rodr{\'{\i}}guez-Gil} {et~al.}(2004){Rodr{\'{\i}}guez-Gil}, {G{\"
  a}nsicke}, {Barwig}, {Hagen}, \& {Engels}}]{rodriguez-giletal04-2}
{Rodr{\'{\i}}guez-Gil}, P., {G{\" a}nsicke}, B.~T., {Barwig}, H., {Hagen},
  H.-J., \& {Engels}, D. 2004, A\&A, 424, 647

\bibitem[{{Rodr{\'{\i}}guez-Gil} {et~al.}(2007){Rodr{\'{\i}}guez-Gil},
  {G{\"a}nsicke}, {Hagen}, {Araujo-Betancor}, {Aungwerojwit}, {Allende Prieto},
  {Boyd}, {Casares}, {Engels}, {Giannakis}, {Harlaftis}, {Kube}, {Lehto},
  {Mart{\'{\i}}nez-Pais}, {Schwarz}, {Skidmore}, {Staude}, \&
  {Torres}}]{rodriguez-giletal07-1}
{Rodr{\'{\i}}guez-Gil}, P., {G{\"a}nsicke}, B.~T., {Hagen}, H.-J., {et~al.}
  2007, MNRAS, 377, 1747

\bibitem[{{Rodr{\'{\i}}guez-Gil} {et~al.}(2005){Rodr{\'{\i}}guez-Gil},
  {G{\"a}nsicke}, {Hagen}, {Nogami}, {Torres}, {Lehto}, {Aungwerojwit},
  {Littlefair}, {Araujo-Betancor}, \& {Engels}}]{rodriguez-giletal05-2}
{Rodr{\'{\i}}guez-Gil}, P., {G{\"a}nsicke}, B.~T., {Hagen}, H.-J., {et~al.}
  2005, A\&A, 440, 701

\bibitem[{{Schenker} \& {King}(2002)}]{schenker+king02-1}
{Schenker}, K. \& {King}, A.~R. 2002, in The Physics of Cataclysmic Variables
  and Related Objects, ed. B.~T. {G\"ansicke}, K.~{Beuermann}, \& K.~{Reinsch}
  (ASP Conf. Ser. 261), 242--251

\bibitem[{{Schenker} {et~al.}(2002){Schenker}, {King}, {Kolb}, {Wynn}, \&
  {Zhang}}]{schenkeretal02-1}
{Schenker}, K., {King}, A.~R., {Kolb}, U., {Wynn}, G.~A., \& {Zhang}, Z. 2002,
  MNRAS, 337, 1105

\bibitem[{{Schlegel} {et~al.}(1998){Schlegel}, {Finkbeiner}, \&
  {Davis}}]{schlegeletal98-1}
{Schlegel}, D.~J., {Finkbeiner}, D.~P., \& {Davis}, M. 1998, apj, 500, 525

\bibitem[{{Schneider} \& {Young}(1980)}]{schneider+young80-2}
{Schneider}, D.~P. \& {Young}, P. 1980, ApJ, 238, 946

\bibitem[{{Schwarzenberg-Czerny}(1989)}]{schwarzenberg-czerny89-1}
{Schwarzenberg-Czerny}, A. 1989, MNRAS, 241, 153

\bibitem[{{Shafter} {et~al.}(1995){Shafter}, {Veal}, \&
  {Robinson}}]{shafteretal95-1}
{Shafter}, A.~W., {Veal}, J.~M., \& {Robinson}, E.~L. 1995, ApJ, 440, 853

\bibitem[{{Smith} \& {Dhillon}(1998)}]{smith+dhillon98-1}
{Smith}, D.~A. \& {Dhillon}, V.~S. 1998, MNRAS, 301, 767

\bibitem[{{Steeghs} \& {Jonker}(2007)}]{steeghs+jonker07-1}
{Steeghs}, D. \& {Jonker}, P.~G. 2007, ApJ Lett., 669, L85

\bibitem[{{Szkody} {et~al.}(2001){Szkody}, {G\"ansicke}, {Fried}, {Heber}, \&
  {Erb}}]{szkodyetal01-1}
{Szkody}, P., {G\"ansicke}, B., {Fried}, R.~E., {Heber}, U., \& {Erb}, D.~K.
  2001, PASP, 113, 1215

\bibitem[{{Thorstensen} \& {Fenton}(2002)}]{thorstensen+fenton02-1}
{Thorstensen}, J.~R. \& {Fenton}, W.~H. 2002, PASP, 114, 74

\bibitem[{{Tonry} \& {Davis}(1979)}]{tonry+davis79-1}
{Tonry}, J. \& {Davis}, M. 1979, AJ, 84, 1511

\end{thebibliography}

\end{document}